# The method of solving initial structure by Seidel aberration theory for extreme ultraviolet lithography objective

## WEI TAN,[1] AND DONGLIN MA[1, *]

*1 School of Optical and Electronic Information and Wuhan National Laboratory of Optoelectronics, Huazhong University of Science and Technology, Wuhan 430074, China*
*madonglin@hust.edu.cn*

**Abstract:** In this paper, a method for solving the initial structure of an off-axis multi-mirror system applied to extreme ultraviolet (EUV) lithography using a paraxial ray-tracing algorithm based on Seidel aberration theory is proposed. By tracing the characteristic rays in the reflection system, the height and paraxial angle on each surface can be obtained, then through the relationship between the Seidel aberration coefficient and these parameters, the initial structure with good aberration performance can be solved. We can obtain different initial structures by adding different initial condition constraints. In this paper, we have solved two different initial structures by assigning different optical powers as different initial structures, and on this basis, we have optimized two off-axis six-mirror systems with numerical aperture (NA) of 0.25. Their wavefront aberration RMS value is about 0.04 wavelength, and the absolute distortion is less than 1.2nm, with good imaging quality. We believe that this method can greatly improve the design efficiency and optimization effect of complex multi-mirror systems.

## 1. Introduction

The resolution of lithography can be determined by the Rayleigh resolution criterion Res=k·λ/NA. Thus reducing the wavelength λ or increasing the numerical aperture (NA) are effective methods to improve the resolution. Extreme ultraviolet (EUV) lithography using 13.5 nm wavelength is the key core of next-generation lithography research and is also a candidate technology for integrated circuit manufacturing to achieve 10nm and below node. As early as 2006, ASML provided the first EUVL industrialization experimental prototype AlphaDemo Tool (ADT), its NA is 0.25, the magnification is 4, and the resolution of 35nm can be achieved [1]. Similarly, the companies such as Nikon and Carl Zeiss have also invested enough research in the EUVL field and has its own extreme ultraviolet lithography system [2,3]. There are many classification methods for existing lithography objective lenses according to different characteristics. It is generally classified according to the number of mirrors, such as six mirrors, eight mirrors and ten mirrors shown in Fig.1[4-6].

In the past few decades, a great deal of research has been done for the design of EUVL objective. At the same time, the design method of the initial structure is also a focus of designers [7]. A good initial structure determines the efficiency and potential of subsequent optimization, which can save a lot of time and experience. However, one of the difficulties in designing an off-axis reflection system is that it is difficult to find the initial structure. Generally, the coaxial reflection system is used as the starting point or the existing off-axis design is used. These methods are cumbersome and time-consuming. A simpler and more convenient method of initial structural design is desired. Fei Liu proposed a grouping design method for an eight-mirror projection objective. By separating the eight-mirror objective into three mirror groups, this method allows designers to calculate the parameters of an eight-mirror objective [8]. Then she extended application range of this method to six-ten mirrors [9]. Based on their research,

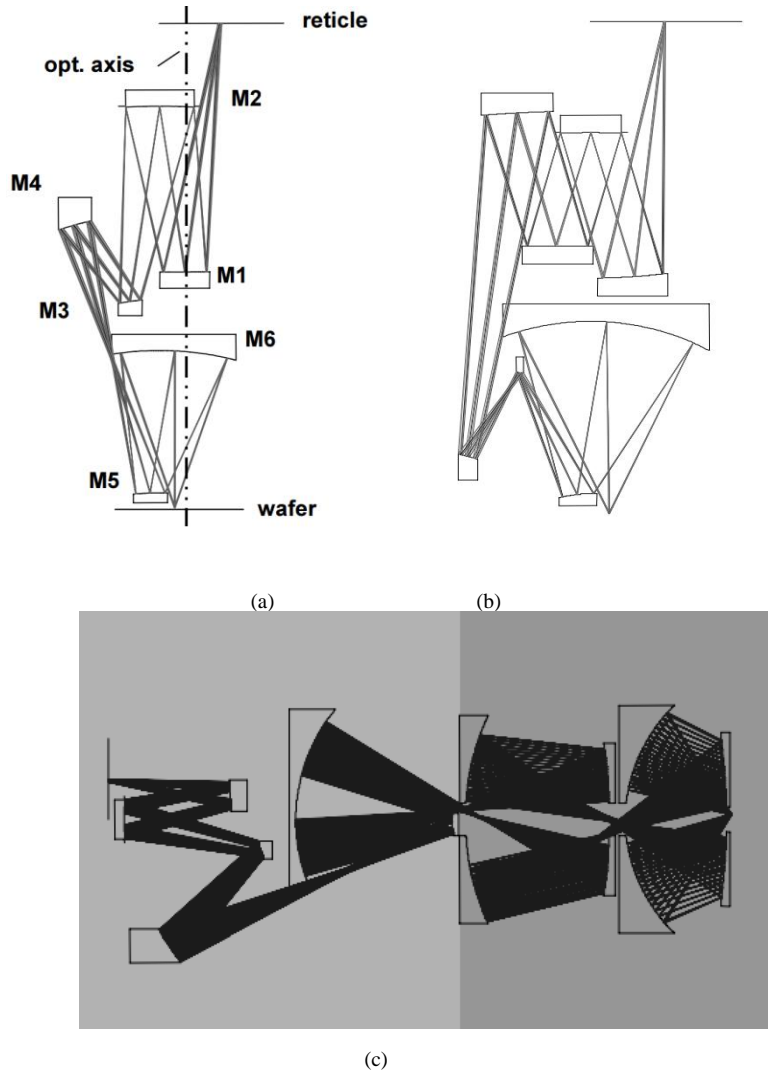

Fig.1 The EUVL objective with different mirrors:(a) six-mirrors(b)eight-mirrors(c)ten-mirrors.

Yan Liu presented a curvatures combination method for an anamorphic EUVL objective. this method achieves an anamorphic magnification initial structure by combining the curvatures of the corresponding surfaces into two coaxial spherical systems [10]. Jun Wang presented the visual generation of the initial construction of EUVL objectives based on the analysis of the valid objectives and grouping strategy which increasing visualization of the design process[11]. These methods usually design some systems with two coaxial spherical surfaces firstly, and then combine each system to form the final initial structure. These methods are cumbersome and do not consider aberration properties.

In this paper, we propose a design method for an off-axis six-mirror EUVL objective based on the coaxial Seidel aberration theory. Firstly, we can obtain the height and paraxial angle on each surface by tracing the characteristic rays in the reflection system, the Seidel aberration of the reflection system can be characterized by these parameters what we get above. Then we can solve the initial structure with no primary aberrations by setting the corresponding Seidel aberration coefficient to zero. Finally, based on the initial structure, a EUVL objective can be quickly optimized with numerical aperture (NA) of 0.25, and wavefront error of 0.04λ.

## 2. INITIAL STRUCTURE DETERMINATION VIA THIRD-ORDER SEIDEL ABERRATION

*a. paraxial ray tracing*

We start the initial design by building a coaxial structure that includes six-mirror spherical surface, as shown in Fig. 2. The aperture stop of system is on the surface M2 which is the second mirror on the system. Characteristic rays are propagating in the reflection system, and the corresponding ray height and paraxial angle are obtained on each surface. The right light propagation angle is necessary to deciding to avoid the occlusion between the surfaces. And the ti mean the distance between the surface Mi and Mi+1, ri denote the curvature radii of the surface Mi. We set transmission media refractive index as ni+1= n′i= -ni, n1=1 in a reflective systerm.

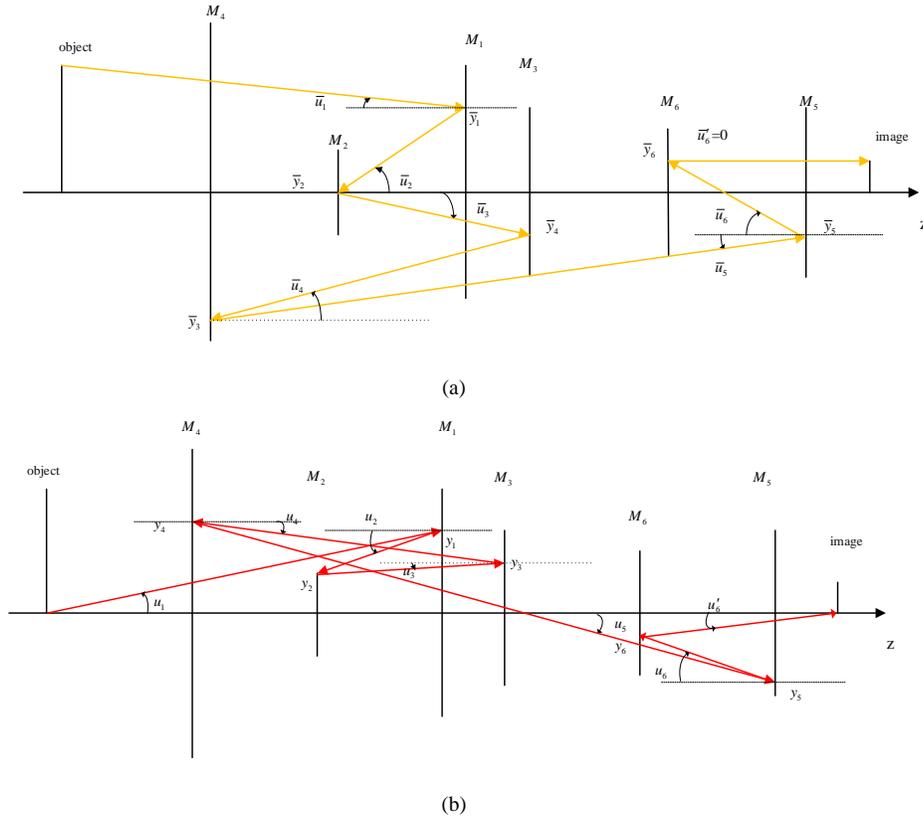

Fig. 2. Ray tracing of the initial coaxial TMA structure: (a) trace of a chief ray and (b) trace of a marginal ray

For the propagation of the marginal ray, as illustrated in Fig.1(b), u is the paraxial ray's incident slope with respect to the optical axis, and u′ is the corresponding exit slope in Fig.1(a). Similarly, yi represent the height of marginal ray on the surface Mi. For the chief ray, the parameter $\bar{u}$, $\bar{u}'$, and $\bar{y}_i$ are same defined.

When the characteristic ray passes through an optical surface, the changing relationship between the ray height and the paraxial angle can be determined by the paraxial ray tracing formulas, which are expressed as

$$n'u' = nu - y\phi$$
$$y' = y + u't'$$
$$\phi = (n' - n) \cdot c \tag{1}$$

When the characteristic ray propagates on multiple optical surfaces, the transfer formulas are expressed as

$$u_{i+1} = u'_i$$
$$y_{i+1} = y'_i$$
$$t_{i+1} = t'_i \tag{2}$$

Where $\Phi_i$ and $c_i$ represent the optical power and the curvature of the surface $M_i$ respectively. For the chief ray, the Eq (1) and Eq (2) can also apply. During ray propagation, there are many optical invariants which can be defined by the paremeters of tracing two paraxial rays (the marginal ray and the chief ray).

$$A = n(yc + u)$$
$$\bar{A} = n(\bar{y}c + \bar{u})$$
$$H = n\bar{u}y - nu\bar{y} \tag{3}$$

Where A represents the Snell invariant of the marginal ray, $\bar{A}$ denotes the Snell invariant of the chief ray, and H is the Lagrange invariant of the system.

According to the Seidel aberration theory, these five monochromatic aberrations, spherical aberration (SI), coma (SII), astigmatism (SIII), Petzval curvature of field (SIV), and distortion (SV), can be expressed by the optical invariants of the rotationally symmetric reflective system, the formulas is

$$S_\mathrm{I} = -\sum A^2 \cdot y \cdot \Delta\left(\frac{u}{n}\right)$$
$$S_\mathrm{II} = -\sum \bar{A} A \cdot y \cdot \Delta\left(\frac{u}{n}\right)$$
$$S_\mathrm{III} = -\sum \bar{A}^2 \cdot y \cdot \Delta\left(\frac{u}{n}\right)$$
$$S_\mathrm{IV} = -\sum H^2 \cdot c \cdot \Delta\left(\frac{1}{n}\right)$$
$$S_\mathrm{V} = -\sum \left\{ \frac{\bar{A}^3}{A} \cdot y \cdot \Delta\left(\frac{u}{n}\right) + \frac{A}{A} \cdot H^2 \cdot c \cdot \Delta\left(\frac{1}{n}\right) \right\} \tag{4}$$

*b. Design of the Initial Structure*

From Eqs. (1)-(4), we know that the Seidel coefficients can be expressed as the functions of the curvature of the surface $c_i$ and the distance between the sufcaces $t_i$ as

$$F = f(c_i, t_i) \qquad (5)$$

Where r and t are the independent variables we need to solve. For an initial structure with no primary aberrations, a specific Seidel coefficient is expected to be zero. We can add initial condition constraints based on design metrics or properties of characteristic rays to solve these equations of aberrations. The chief ray angle on mask should be less than 6° avoiding the shadow effects on mask means $\bar{u}_1 \leq 6°$, and image telecentricity of lithography system means $\bar{u}'_6 = 0$. There are many other initial conditions to limit the initial structure such as the number aperture, the magnification and field of view etc.

We can obtain different initial structures by adding different initial condition constraints. There are two kinds of optical power distribution commonly used in EUV lithography objective lens: PNNPNP and PPNPNP. By assigning different optical powers as different initial structures, we have solved two different initial structures showed in fig 2(a) and fig.3(a). From the fig 2(b) and fig.3(b), wo can get that the initial structures have the good image quality when it is in low NA. This means initial structures can be a good basis for a design starting point.

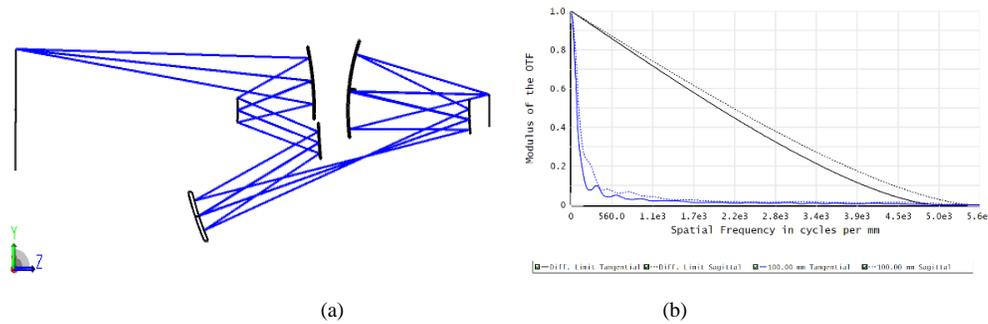

(a)  (b)

Fig.3 initial structures with the optical powers of PNNPNP: (a) Layout (b)MTF

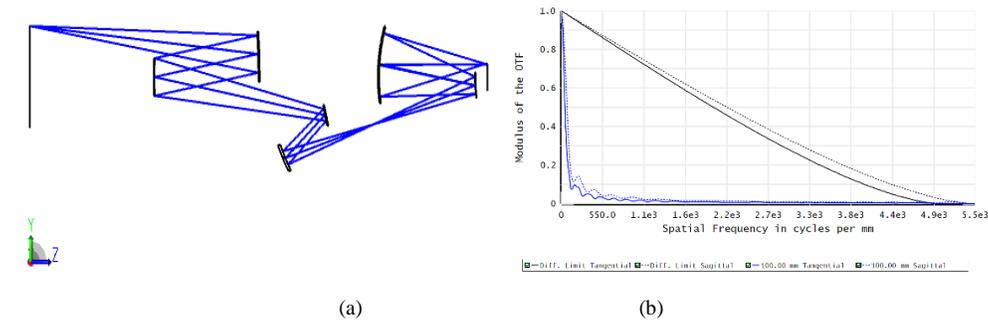

(a)  (b)

Fig. 4. initial structures with the optical powers of PPNPNP: (a) Layout (b)MTF

## 3. DESIGN RESULTS AND PERFORMANCE AND TOLERANCE ANALYSIS

### *a. design results*

After obtaining the initial structure solved above, a further optimization process will be applied to get a EUV lithography objectives with good imaging quality. First, we will optimize an objective lens system with a small NA, using a small number of variables to make the wavefront aberration and distortion meet the requirements, increase the optimization variables while increasing the NA, and repeat the above process until the NA meets the requirements. Finally, we obtained two off-axis six-mirror EUV lithography objective with different optical power distributions.

For the EUV lithography objectives with PNNPNP optical power and NA 0.25, the final layout is shown in Fig.5(a), and the finally optical parameters of the objective are described in Table 1.

Table 1. optical characteristics of EUVL objective with PNNPNP optical power

| Parameters | Specifications |
| --- | --- |
| Wavelength | 13.5nm |
| numerical aperture | 0.25 |
| field of view | 26mm×1.5mm |
| magnification | 1/4 |
| wavefront error | <0.52nm(0.4λ) |
| max distortion | 1.2nm |
| Chief ray angle on the mask | 5.06° |
| image telecentricity | 0.01° |

The final off-axis six-mirror systerm has an NA of 0.25 and a 26mm × 1.5mm arc field of view. And the chief ray angle on the mask is 5.06°, image telecentricity is 0.02°. The frequency Optical Modulation Transfer Function (MTF) plot is exhibited in Fig.5(b, the image quality almost reaches the diffraction limitation. Form the Fig.5(c) and Fig.5(d), wo can know that the maximum RMS wavefront aberration of the objective in the full field of view is 0.039λ, less than 0.53nm, and the maximum distortion is 1.2nm.

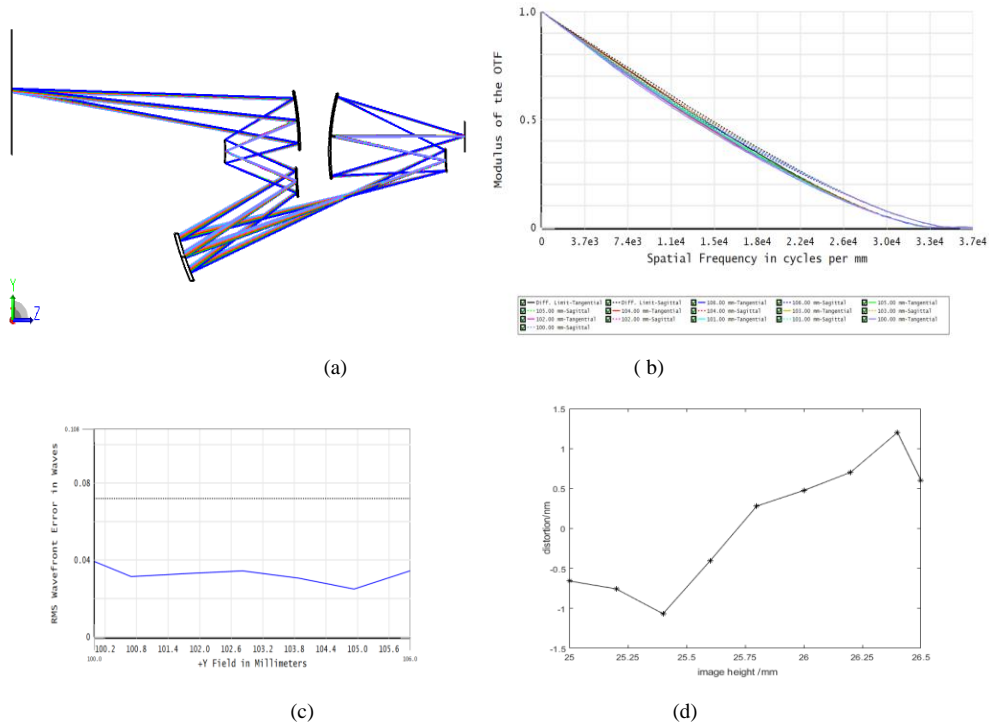

Fig.5 Image quality of the system with PNNPNP optical power: (a) layout (b) MTF. (c) RMS wavefront error. (d) distortion.

For the EUV lithography objectives with PPNPNP optical power and NA 0.25, the final layout is shown in Fig.6(a), and the finally optical parameters of the objective are described in Table 2.

Table 2. optical characteristics of EUVL objective with PPNPNP optical power

| Parameters | Specifications |
| --- | --- |
| Wavelength | 13.5nm |
| numerical aperture | 0.25 |
| field of view | 26mm×1.5mm |
| magnification | 1/4 |
| wavefront error | ＜0.58nm(0.43λ) |
| max distortion | 0.8nm |
| Chief ray angle on the mask | 5.59° |
| image telecentricity | 0.02° |

The structure 2 has different optical characteristics with the structure 1. The chief ray angle on the mask is 5.59°, image telecentricity is 0.02°. The (MTF) plot is exhibited in Fig.6(b), and the image quality also approaches the diffraction limitation. The maximum RMS wavefront aberration of the objective in the full field of view is 0.043λ, less than 0.58nm as shown in Fig.6(c), and the maximum distortion is 0.8nm as shown in Fig.6(d).

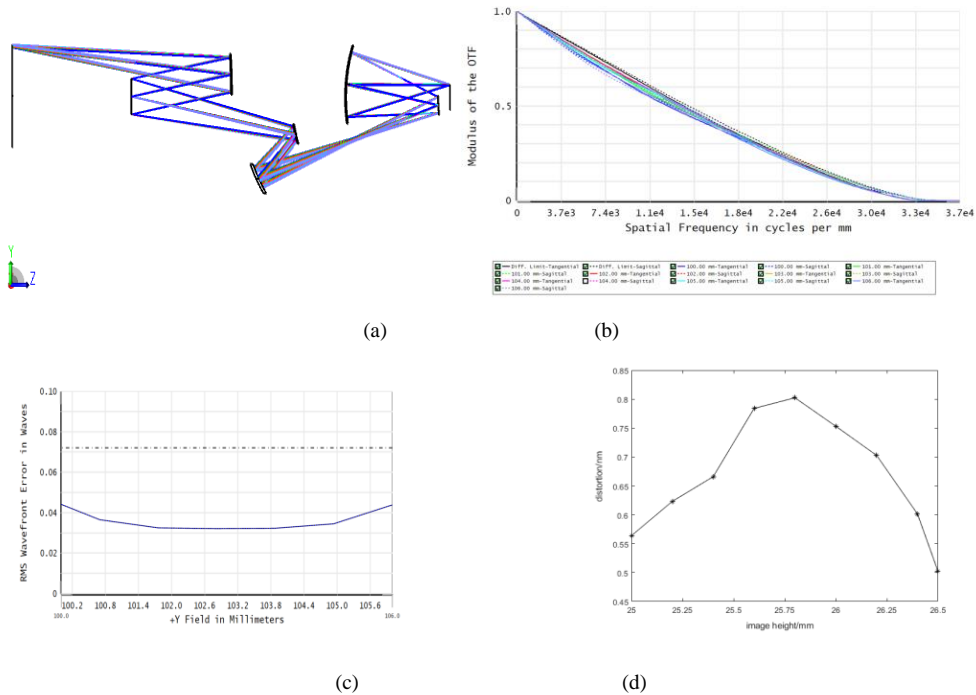

Fig.6 Image quality of the system with PPNPNP optical power: (a) layout (b) MTF. (c) RMS wavefront error. (d) distortion.

We obtained different initial structures with different optical powers, and on this basis, we optimized two different final structures, which wavefront error and distortion can all meet the requirements. This demonstrate the feasibility and superiority of the initial structure by the method.

*b. tolerance analysis*

For the EUV lithography objective lens based on the off-axis reflection system, because of its extremely high image quality requirements, there is also s a very high limit to the processing accuracy, which is more difficult to manufacture and assemble than the ordinary coaxial system. We conduct a simple tolerance analysis of the EUVL objectives with PPNPNP optical power. The tolerance distribution of each component is listed in Table 3.

**Table 3. Tolerance Settings in Zemax**

| surface | Radius | Distance | Decenter in x | Decenter in y | tilt in x | tilt in y |
| --- | --- | --- | --- | --- | --- | --- |
| M1 | ±1.0um | ±0.1um | ±0.1um | ±0.1um | ±0.36″ | ±0.36″ |
| M2 | ±1.0um | ±0.1um | ±1.0um | ±0.1um | ±0.72″ | ±0.72″ |
| M3 | ±0.1um | ±0.01um | ±0.01um | ±0.01um | ±0.36″ | ±0.36″ |
| M4 | ±0.01um | ±0.01um | ±0.01um | ±0.01um | ±0.036″ | ±0.036″ |
| M5 | ±0.1um | ±0.1um | ±0.1um | ±0.01um | ±0.36″ | ±0.36″ |
| M6 | ±0.1um | ±0.1um | ±0.05um | ±0.05um | ±0.36″ | ±0.36″ |

We select the average diffraction modulation transfer function (MTF) at 15000 lp/mm as the criteria to evaluate the optical performance, then performed 5000 simulations using Monte Carlo sensitivity analysis to predict the optical performance. The overall analysis results are shown in table 4. As shown in table 3 and table 4, a 98% MTF is larger than 0.445 at 15000lp/mm, the surface M4 is the most sensitive to aberrations, , and has the tightest tolerance, so M4 has the greatest impact on system manufacturing and can be used as the benchmark for assembly.

**Table 4. Monte Carlo Tolerance Analysis Probability Results of MTF at 15000lp/mm**

| Proportion | 98% | 90% | 80% | 50% |
| --- | --- | --- | --- | --- |
| MTF | 0.4450 | 0.4476 | 0.4492 | 0.4510 |

## 4. CONCLUSION

In this paper, two different off-axis six-mirror EUVL objective based on Seidel aberration theory has been successfully designed. The Seidel aberration in the coaxial reflection system can be characterized by the paraxial angle and height of characteristic rays on each optical surface using paraxial ray tracing. And we construct the functional relationship between the Seidel aberration and the curvature of the optical surface and the optical distance. Through different initial conditions, different initial structures are solved. On this basis, two EUVL objective lenses with 0.25 NA and 1/4 magnification are obtained by optimization. The wave aberration of the two structures is less than 0.6nm, and the distortion is also around 1nm, which demonstrates potential of this design method. And, a brief tolerance analysis is provided to demonstrate the instrumentation feasibility of the EUVL objective. In this design, the mirror has an aspheric surface with 16th-order coefficients. If more complex surfaces, such as extended polynomials surface, are used, a larger NA and better image quality can theoretically be achieved, which will be carried out in future studies.

**Disclosures.**

The authors declare no conflicts of interest.

**Data availability**

Data underlying the results presented in this paper are not publicly available at this time but may be obtained from the authors upon reasonable request